\documentclass[pra,twocolumn,showpacs,preprintnumbers,amsmath,amssymb]{revtex4-1}

\usepackage{graphicx,natbib}
\usepackage{dcolumn}
\usepackage{bm}
\usepackage{xcolor}

\begin{document}

\title{Gravitational Vortex Mass in a Superfluid}

\author{Tapio Simula}
\affiliation{Centre for Quantum and Optical Science, Swinburne University of Technology, Melbourne 3122, Australia}

\begin{abstract}
{We consider superfluid hydrodynamics of two-dimensional Bose--Einstein condensates. Interpreting the curvature of the macroscopic condensate wavefunction as an effective gravity in such a superfluid universe, we argue for a superfluid equivalence principle---that the gravitational mass of a quantised vortex should be equal to the inertial vortex mass. In this model, gravity and electromagnetism have the same origin and are emergent properties of the superfluid universe, which itself emerges from the underlying collective structure of more elementary particles, such as atoms. The Bose--Einstein condensate is identified as the elusive dark matter of the superfluid universe with vortices and phonons, respectively, corresponding to massive charged particles and massless photons. Implications of this cosmological picture of superfluids to the physics of dense vortex matter are considered.
}
\end{abstract}

\maketitle

\section{Introduction}

The apparent equivalence of the inertial mass $m_{\rm i}$ and the gravitational mass $m_{\rm g}$ has puzzled scientists at least since the times of Newton \cite{newton1687philosophiae}. Einstein's equivalence principle is a cornerstone of general relativity that necessitates this equality, albeit provides no explanation for its origin \cite{lorentz1952principle}. More recently, the repeated attempts to unify quantum mechanics and general relativity, together with the growing mystery of the dark matter paradigm \cite{Zwicky1933a,Rubin_2000,Peebles2017a}, are calling for revision to our understanding  of the nature of gravity and the fabric of spacetime. 

Here we bypass such grand challenges and translate the question of the equivalence between inertial and gravitational masses to a superfluid toy universe, whose properties are based on firm theoretical and experimental foundations. The full superfluid spacetime is 2+1 dimensional whereas the spacetime of the particles (vortices) of this theory is 1+1 dimensional. A remarkable property of three dimensional spacetimes is that they may allow for a variety of well founded formulations of quantum gravity \cite{Witten2007a}. Further to this, it has been suggested that Einstein's field equations, and thereby gravity, could emerge as a consequence of quantum fluctuations of a regular quantum field theory over a background spacetime metric \cite{Sakharov1967a,misner1973gravitation}. Meanwhile, a correspondence between two-dimensional superfluid hydrodynamics and relativistic electrodynamics is  established \cite{Popov1973a,GriffinPRB,Klein2014a} and underpins the concept of the inertial mass of a vortex \cite{Popov1973a,volovik2003universe,Thouless2007a,Simula2018a,Simula2019book}. In this work, we draw inspiration from these theoretical considerations with a specific focus on aiming to investigate the equivalence principle and the gravitational mass of a quantised vortex within the context of such a emergent superfluid universe.

In hydrodynamic theory of fluids the vorticity ${\boldsymbol \omega} = \nabla\times {\boldsymbol v}$, the curl of the velocity potential, plays a pivotal role. It is already at this elementary level that the connection between hydrodynamics of fluids and classical electromagnetic theory, as quantified by Maxwell's equations, seems to appear, since the magnetic field is equal to the vorticity of the magnetic vector potential. This connection presumably prompted Maxwell to contemplate the hypothesis of molecular vortices and to state that ``\emph{under the action of magnetic forces something belonging to the same mathematical class as angular velocity$\ldots$forms a part of the phenomenon}" \cite{Maxwell}. Presently, it is thought that the unified electromagnetic field describes all of the classical electromagnetic phenomena, including the propagation of light. Quantum electrodynamics further explains how charged particles may be spawned as excitations of the electromagnetic field \cite{Dyson1965a}. However, neither Maxwell's electrodynamics or quantum electrodynamics are able to shed any light on the nature of the `substrate' (infamously known as the aether) of the electromagnetic field.

Two-dimensional (plus one time dimension) superfluids provide a mathematically appealing analogue to relativistic electrodynamics \cite{Popov1973a,GriffinPRB,Klein2014a}. In such a superfluid universe, the electric field is associated with the superflow as determined by the gradient of the spatial phase of the condensate order parameter of the superfluid. The magnetic field may be associated with the rate of change of the dynamic phase of the condensate, and the quantised vortices (more precisely the kelvons which are quasiparticles associated with the vortex) correspond to the massive charged particles, analogous to the electrons. In contrast to electromagnetic fields of our Universe, the `substrate' of the electromagnetic fields of this superfluid universe correspond to a well defined entity---a Bose--Einstein condensate comprised of the underlying `trans-Planckian' constituent particles such as rubidium atoms \cite{volovik2003universe}. 

Once the Bose--Einstein condensate forms, the quasiparticles of the condensate are elevated to the status of the elementary particles of the superfluid universe and its vacuum, the substrate for all fields, is the condensate itself. In this sense, the whole superfluid universe together with all of the fundamental forces are emergent. The trans-Planckian atoms respect Galilean invariance but the quasiparticles of the superfluid with the linear phonon quasiparticle dispersion relation allows for an interpretation in terms of an acoustic metric with an effective Lorentz invariance \cite{Visser2005a}. In this picture, the quantum field theory of the normal state atoms realise the grand unified theory (GUT) of the superfluid universe and the Bose--Einstein condensate corresponds to a low energy state that emerges via a spontaneous symmetry breaking mechanism as the system cools. The superfluid universe features a peculiar anti-GUT property whereby new effective symmetries emerge in the `low energy corner' of the quasiparticles \cite{volovik2003universe}, as quantified by the Bogoliubov dispersion relation
\begin{equation}
E(p) = \sqrt{(pc_s)^2 + \left(\frac{p^2}{2m}\right)^2},
\end{equation}
where $c_s$ is the speed of sound and $p$ is the quasiparticle momentum. These are the `relativistic' Bogoliubov phonons with an acoustic metric associated with the linear dispersion relation at low momenta $p\to 0$ that results in the emergent Lorentz invariance. In contrast, the effective Lorentz invariance violating term, $\propto p^4$ in the square root, results in a quadratic dispersion relation for high momenta and at high temperatures. 

The true zero mode of this system, the Nambu--Goldstone boson, is the vacuum (the condensate) of this theory.  The topological excitations (quantised vortices) with angular momentum quantum number $\ell=-1$ kelvons are the charged particles of this superfluid universe. Their dispersion relation $\omega = \omega_k+k^2\ln(1/k)$ is approximately linear at low momenta and may be viewed as the effective relativistic particles of the theory \cite{Popov1973a}. The kelvon based inertial mass of a vortex, which is the electron of the superfluid universe, is \cite{Simula2018a}
\begin{equation}
m^{v}_{\rm i}=  \frac{2\pi\hbar n}{ \omega_k},
\label{restmass}
\end{equation}
where $\omega_k$ is the zero-point frequency of the kelvon and $n$ is the two-dimensional background condensate particle density. The sound waves in the superfluid are the photons of the superfluid universe and quantum turbulence gives rise to an emergent gravitational field via the resulting non-vanishing quantum pressure field, as discussed in detail later.

In Section II, we associate each term in the generalised Gross--Pitaevskii energy functional with fields of the superfluid universe. Sections III and IV discuss the emergence of electromagnetism and gravity, respectively. In Section V we argue for the vortex correspondence principle proposing the equality of the gravitational and inertial vortex mass. In a 2+1 dimensional superfluid the motion of quantised vortices may be modelled in terms of Hamilton's phase-space equations for a one dimensional massive particle. The resulting vortex-particle duality is considered in Section VI. In Section VII we consider quantum Hall physics in the superfluid universe. Concluding remarks are provided in Section VIII.

\section{A (2+1)-dimensional superfluid universe}

We consider a two-dimensional (plus one time dimension) superfluid universe governed by the order parameter $\Phi({\bf r},t)$ normalised to the atom number $N_a=\int|\Phi|^2d{\bf r}^2$ and the usual Gross--Pitaevskii energy functional \cite{pethick2002bose,pita2003bose} 
\begin{equation}
 \mathcal{E} =\int \left ( \frac{\hbar^2}{2m}|\nabla\Phi |^2   +  \frac{c_0}{2}|\Phi |^4+2c_0\tilde{n}|\Phi |^2 -\mu_{\rm DE} |\Phi |^2\right )d{\bf r}^2,
\label{GPEegy}
\end{equation}
where $m$ is the mass of the `trans-Plankian' particles (e.g. atoms) and $c_0$ is the coupling constant that relates the condensate density to the energy per particle (chemical potential) $\mu_{\rm DE}$ of the superfluid vacuum.  The evolution of this superfluid universe is determined by the generalized Gross-Pitaevskii equation
\begin{equation}
i\hbar\partial_t \Phi({\bf r}, t) =  \mathcal{H}  \Phi({\bf r}, t),
\label{GPEtd}
\end{equation}
where the Hamiltonian is defined by
\begin{equation}
\mathcal{H}=\left( -\frac{\hbar^2}{2m}\nabla^2   +  c_0n({\bf r},t) +  2c_0\tilde{n}({\bf r},t) -\mu_{\rm DE}  \right),
\label{GPham}
\end{equation}
$n({\bf r}, t)= |\Phi({\bf r}, t)|^2$ and $\tilde{n}({\bf r}, t)$ is the particle density of the fluid not included in the condensate. Using Madelung transformation \cite{Madelung1927a} $\Phi({\bf r}, t)= |\Phi({\bf r}, t)|e^{iS({\bf r}, t)}$, in the case of static thermal cloud, Eq.~(\ref{GPEtd}) may be expressed equivalently in its hydrodynamic form in terms of continuity equation
\begin{equation}
\frac{\partial n}{\partial t} = -\nabla\cdot \left(  n {\boldsymbol v}_s \right)
\label{GPEcont}
\end{equation}
and an Euler-like equation
\begin{equation}
-\hbar\frac{\partial S}{\partial t} = -\frac{\hbar^2}{2m|\Phi|}\nabla^2|\Phi| +  \frac{m}{2}{v}_s^2 +  c_0|\Phi|^2 +  2c_0\tilde{n} -\mu_{\rm DE},
\label{GPEeuler}
\end{equation}
where ${\boldsymbol v}_s=\hbar\nabla S/m$ is the superfluid velocity and $S$ is the scalar valued phase function of the condensate \cite{pethick2002bose,pita2003bose}.

To draw a distinction between the standard terminology used in cold atom physics and that of the superfluid universe, we change the notation in Eq.~(\ref{GPEegy}) expressing it as
\begin{equation}
 \mathcal{E} = {\rm GEM} + \int  \left [ \left(\frac{c_0}{2}\Psi^2_{\rm DM}+2c_0\Psi^2_{\rm NM}-\mu_{\rm DE} \right)\Psi^2_{\rm DM} \right ]d{\bf r}^2,\notag
\label{GP22}
\end{equation}
where GEM, defined later, stands for gravity and electromagnetism, the dark matter density $\Psi^2_{\rm DM}=|\Phi({\bf r},t)|^2$ and the normal matter density $\Psi^2_{\rm NM}=\tilde{n}({\bf r},t)$. 
In equilibrium, the normal matter
\begin{equation}
\tilde{n}({\bf r}) = \sum_{q} \big\{ f(T,E_q) [|u_q({\bf r})|^2 + |v_q({\bf r})|^2 ] + |v_q({\bf r})|^2 \big\}
\label{rho}
\end{equation}
exists in the form of Bogoliubov quasiparticles with energies $E_q$ and quasiparticle amplitudes $u_q$ and $v_q$. The Bose--Einstein distribution $f(T,E_q)$ determines the dependence of the normal matter density on temperature $T$ of the atoms. Two main types of luminous quasiparticle matter in this theory are phonons (massless particles) and kelvons (charged, massive particles), the latter being confined to and carried along by quantised vortices. The quasiparticles are oblivious to the existence of the trans-Planckian world of atoms out of which the condensate and thereby the whole superfluid universe emerged. Although the whole superfluid universe including gravity and electrodynamics is emergent, the rules of quantum mechanics are nevertheless inherited by the quasiparticles of the superfluid from the laws governing the trans-Planckian world of true atoms.

Next we re-employ the Madelung transformation to split the GEM into a part involving supercurrents and another one accounting for the effects of quantum pressure. This yields
\begin{equation}
{\rm GEM}  =\int \frac{\hbar^2}{2m}|\nabla S({\bf r})|^2|\Phi({\bf r})|^2 +\frac{\hbar^2}{2m}\left(\nabla\left|\Phi({\bf r})\right| \right)^2   d{\bf r}^2,
\label{GP3}
\end{equation}
which we may also express as
\begin{equation}
{\rm GEM} =\int \Psi^2_{\rm EM}\Psi^2_{\rm DM} d{\bf r}^2+\int\Psi^2_{\rm G} \Psi^2_{\rm DM} d{\bf r}^2.
\label{GP4}
\end{equation}
The electromagnetic interaction $\Psi^2_{\rm EM}$ emerges due to the motion (flow of the true atoms) of the vacuum and the gravitational interaction $\Psi^2_{\rm G}$ emerges due to the curvature (variation in the density of the true atoms) of the superfluid spacetime. The effective interaction between gravitational and electromagnetic fields is mediated by the dark matter $\Psi^2_{\rm DM}$. We will first consider the $\Psi^2_{\rm EM}$ term.

\section{Emergent electromagnetism}

Let us first consider the superfluid density $n=n_0+\delta n$ to be uniform (yet, oxymoronically allowing infinitesimal density fluctuations $\delta n$) and the trans-Plankian atoms to be confined to a quasi-2D regime such that the embedding space is three-dimensional but the vortex dynamics is planar (two-dimensional). This is a typical setting for instance in experimental studies on two-dimensional quantum turbulence in Bose--Einstein condensates \cite{Gauthier2019a,Johnstone2019a}.
The term
\begin{equation}
{\rm GEM}_{\rm EM}  =\int \Psi^2_{\rm EM}|\Phi({\bf r})|^2  d{\bf r}^2,
\label{GP}
\end{equation}
corresponds to the electromagnetic energy of the superfluid universe and the classical electrodynamics are obtained by considering the superfluid hydrodynamics within the uniform condensate density approximation such that the continuity equation (\ref{GPEcont}) becomes
\begin{equation}
\frac{\partial n_0}{\partial t} = -n_0\left( \nabla_\perp\cdot {\boldsymbol v}_s \right),
\label{continuity}
\end{equation}
where we have introduced the subscript $\perp$ to remind us of the fact that spatial gradients only exist in the two-dimensional plane, and the equation for the phase evolution (\ref{GPEeuler}) may be approximated by
\begin{equation}
-\hbar\frac{\partial S}{\partial t} =    \frac{1}{2}m{v}_s^2 +  c_0n_0  + 2c_0\tilde{n} - \mu_{\rm DE}.
\label{GPEuler2}
\end{equation}
The superfluid electric and magnetic fields 
\begin{equation}
{\bf E}_{\rm sf}=mn_0{\boldsymbol v}_{\rm s}\times {\bf e}_z 
\hspace{5mm}
{\rm and}
\hspace{5mm}
{\bf B}_{\rm sf}= \frac{\hbar m}{c_0}\frac{\partial S}{\partial t} {\bf e}_z,
\label{EB}
\end{equation}
where ${\bf e}_z$ is the unit vector normal to the condensate plane, correspond to, respectively, spatial and temporal gradients of the condensate phase. Defining the magnetic field to be proportional to the phase change $\partial_tS$, rather than the condensate density $n_0$, ensures that the mean value of the magnetic field of the vacuum (ground state condensate) vanishes since then $\partial_t S=c_0n_0-\mu_{\rm DE}=0$.
The vortex current
\[{\bf j}_v = \rho_v{\boldsymbol v}_v, \] 
where ${\boldsymbol v}_v$ is the velocity field of the vortex phase singularities, is expressed in terms of the vortex density
\begin{align}
\rho_v &=  (\nabla_\perp\times{\boldsymbol v}_{\rm s}) \cdot {\bf e}_z.
\end{align}

The superfluid vacuum constants are
\begin{equation}
\epsilon_v=\frac{1}{mn_0}\hspace{10mm}{ \rm and} \hspace{10mm} \mu_v=\frac{m^2}{c_0}
 \end{equation}
such that the speed of sound is
\begin{equation}
c_s = \sqrt{\frac{c_0n_0}{m}}=\frac{1}{\sqrt{\mu_v\epsilon_v}}.
\end{equation}
With these definitions, all of the classical electrodynamic theory for the superfluid universe can be derived starting from the generalized Gross--Pitaevskii energy functional.

\subsection{Gauss-like E law}
The Gauss-like law \begin{equation}
\nabla_\perp\cdot {\bf E}_{\rm sf} = \frac{\rho_v}{\epsilon_v}
\label{GPgausse}
\end{equation}
states that the vortex `charges' are the sources of the electric field, and is merely a re-statement of the quantization of circulation $\oint {\boldsymbol v}_s\cdot {\rm d}{\bf l} = \kappa w$, where $w$ is an integer winding number and $\kappa=2\pi\hbar/m$ is the quantum of circulation. Using Eq.~(\ref{EB}) the divergence of the electric field is
\begin{align}
\nabla_\perp\cdot {\bf E}_{\rm sf} &=mn_0\nabla_\perp\cdot({\boldsymbol v}_{\rm s}\times {\bf e}_z) \notag\\
&=mn_0 \nabla_\perp\cdot(v_y {\bf e}_x - v_x{\bf e}_y)\notag\\
&=mn_0 (\partial_x v_y - \partial_y v_x).
\end{align}
The vortex density 
\begin{align}
\rho_v &=  \sum_{i=1}^{N_v} w_i\kappa \delta(r-r_i)\notag\\
&=\boldsymbol\omega\cdot {\bf e}_z =(\nabla_\perp\times{\boldsymbol v}_{\rm s}) \cdot {\bf e}_z \notag\\
& =(\partial_xv_y - \partial_y v_x ),
\end{align}
where $N_v$ is the total number of vortices in the system and $w_i$ is the integer winding number of the $i$th vortex, is equal to the divergence of the electric field divided by the permittivity of the vacuum, as stipulated by the Gauss's law. In the continuum limit the Feynman criterion for the areal vortex density yields
\begin{equation}
\rho_v =\kappa n_v = \kappa \frac{\Omega_{\rm rot} m}{\hbar\pi} = 2\Omega_{\rm rot},
\label{feyn}
\end{equation}
which states that the magnitude of the vorticity of a rigidly rotating body equals twice its angular rotation frequency $\Omega_{\rm rot}$.

\subsection{Gauss-like B law}

The Gauss-like law for the magnetic field
\begin{equation}
\nabla_\perp\cdot {\bf B}_{\rm sf}=\nabla_\perp\cdot [B(x,y){\bf e}_z] = 0
\label{GPgaussb}
\end{equation}
is trivially satisfied because ${\bf B}$ has only one component and it is orthogonal to the $x-y$ plane. In words, this superfluid universe has no monopoles of magnetic kind. 

\subsection{Faraday-like law}

The law of induced electric fields due to changing magnetic field
\begin{equation}
\nabla_\perp\times {\bf E}_{\rm sf}  = -\frac{\partial {\bf B}_{\rm sf}}{\partial t}
\label{GPfara}
\end{equation}
may be derived using the continuity equation for the superflow of atoms. The curl of the electric field is
\begin{equation}
\nabla_\perp\times {\bf E}_{\rm sf}=mn_0\nabla_\perp\times({\boldsymbol v}_s\times {\bf e}_z) =-mn_0(\nabla_\perp\cdot{\boldsymbol v}_s){\bf e}_z,  
\end{equation}
where the second equality follows from a vector identity.
The negative of the time derivative of the magnetic field
\begin{equation}
-\frac{\partial {\bf B}_{\rm sf}}{\partial t}= -\frac{\hbar m}{c_0} \frac{\partial}{\partial t} \left( \frac{\partial S}{\partial t}\right) {\bf e}_z
=-mn_0(\nabla_\perp\cdot{\boldsymbol v}_s){\bf e}_z
\end{equation}
is thus equal to the curl of the electric field. This can be shown by differentiating the Euler-like equation (\ref{GPEuler2}) to yield
\begin{equation}
 -\hbar \frac{\partial}{\partial t} \left( \frac{\partial S}{\partial t}\right) =  \frac{\partial}{\partial t} \left(\frac{1}{2}m{v}_s^2 + c_0n_0 +  2c_0\tilde{n} - \mu_{\rm DE}\right).
\label{Euler}
\end{equation}
Under the assumption that the kinetic energy per particle of the flow is conserved, it follows that for a constant $\tilde n$ and $\mu_{\rm DE}$
\begin{equation}
-\frac{\hbar m}{c_0} \frac{\partial}{\partial t} \left( \frac{\partial S}{\partial t}\right) =  m\frac{\partial n_0}{\partial t}  =-mn_0\nabla_\perp\cdot{\boldsymbol v}_s,  
\label{eulerlike2}
\end{equation}
where the second equality is just the continuity equation.

\subsection{Ampere--Maxwell-like law}

The law of induced magnetic fields due to electric current or changing electric field is
\begin{equation}
\nabla_\perp\times {\bf B}_{\rm sf}=\mu_v{\bf j}_v + \mu_v\epsilon_v \frac{\partial {\bf E}_{\rm sf}}{\partial t},
\label{GPmax}
\end{equation}
which is consistent with the vortex current continuity equation $\nabla \cdot{\bf j}_v+\partial _t \rho_v=0$.
As in classical electrodynamics, the charges and currents must be explicitly introduced while in the full theory they emerge naturally as excitations of the superfluid. Once the charges have been introduced, Eq.~(\ref{GPmax}) may be derived by considering a transformation to a reference frame moving at a local vortex velocity
\begin{equation}
i\hbar\partial_t \Phi_v({\bf r}, t) = [i\hbar\partial_t - {\bf J}_v\cdot {\bf A}] \Phi({\bf r}, t).
\label{transform}
\end{equation}
For the case of a uniformly rotating vortex lattice with an approximation that the mean vortex velocity would be half the local superfluid velocity, Eq.~(\ref{transform}) reduces to the usual transformation to a rigidly rotating frame
\begin{equation}
{\bf J}_v\cdot{\bf A} = {\boldsymbol \Omega}_{\rm rot}\cdot ({\bf r}\times {\bf p})=-m{\bf j}_v\cdot{\bf r}\times {\bf e}_z.
\end{equation}

Since for a generic scalar function $A(r)$
\begin{equation}
\nabla_\perp \times [ A(r){\bf e}_z] = [\nabla_\perp A(r)]\times {\bf e}_z,
\label{scalaride}
\end{equation}
the curl of the superfluid magnetic field is
\begin{align}
&\nabla_\perp\times {\bf B}_{\rm sf}=\nabla_\perp\times \left\{\frac{\hbar m}{c_0}\left(\frac{\partial S}{\partial t} -{\bf J}_v\cdot{\bf A}/\hbar  \right){\bf e}_z\right\} \notag\\
&= \frac{\hbar m}{c_0}\nabla_\perp\times \left[\left(\frac{\partial S}{\partial t} + \frac{m}{\hbar} {\bf j}_v \cdot {\bf r}\times {\bf e}_z  \right){\bf e}_z\right] \notag\\
&= \frac{\hbar m}{c_0}\nabla_\perp  \left(\frac{\partial S}{\partial t} \right) \times {\bf e}_z        +\frac{m^2}{c_0}\nabla_\perp\left( {\bf j}_v \cdot {\bf r}\times {\bf e}_z  \right) \times {\bf e}_z  . 
\label{GPE}
\end{align}
The first term on the last row is equal to
\begin{align}
&\mu_v\epsilon_v\frac{\partial {\bf E}_{\rm sf}}{\partial t}=\mu_v\frac{\partial {\boldsymbol v}_{\rm s}}{\partial t}\times {\bf e}_z =\frac{\hbar m}{c_0}\nabla_\perp\left( \frac{\partial {S}}{\partial t} \right)\times {\bf e}_z
\end{align}
and the vortex current 
\begin{equation}
\mu_v{\bf j}_v = \frac{m^2}{c_0}\nabla_\perp\left( {\bf j}_v \cdot {\bf r}\times {\bf e}_z  \right) \times {\bf e}_z 
\end{equation}
results from the fact that the gradient operator does not act on the vortex coordinates ${\boldsymbol r_v}$. 
 
\subsection{Lorentz-like force law}

The exact equation of motion for a vortex is \cite{Groszek2018a}
\begin{equation}
{\boldsymbol v}_v = {\boldsymbol v}_S+ {\boldsymbol v}_n,
\label{vortexeqn}
\end{equation}
where ${\boldsymbol v}_v$ is the vortex velocity and 
\begin{equation}
{\boldsymbol v}_S=\hbar\nabla_\perp S(r)/m|_{r_v} 
\end{equation}
is the background condensate phase gradient (electric field) at the location of the vortex after the vortex self-induction velocity field is removed. The velocity component 
\begin{equation}
{\boldsymbol v}_n = - \frac{\hbar  {\bf e}_z \times \nabla e({\bf r})}{2m e({\bf r})} \bigg |_{r_v} 
\end{equation}
due to the background condensate density gradient (gravity) is expressed in terms of the function $e(r)$ defined in the vicinity of the vortex core by $n(r)=e(r) (r-r_v)^2$. Equation~(\ref{vortexeqn}) can be cast as a force equation 
\begin{align}
 mn_0{\boldsymbol \kappa} \times{\boldsymbol v}_n = -mn_0 {\boldsymbol \kappa}\times({\boldsymbol v}_S -{\boldsymbol v}_v). 
\label{Magnus2}
\end{align} 
Direct substitution of Eq.~(\ref{EB}) shows that Eq.~(\ref{Magnus2}) is just the superfluid Lorentz force law 
\begin{align}
{\bf F}_{\rm Lorentz}&=\kappa {\bf E}_{\rm sf} +\kappa{\boldsymbol v}_v \times {\bf B}_{\rm sf},
\label{Lorentz}
\end{align} 
where ${\bf B}_{\rm sf}\approx -mn_0{\bf e}_z$ since the dynamical phase evolution at the location of the vortex phase singularity where $n|_{r_v}=0$ is $\partial S/\partial t|_{r_v}\approx- c_0n_0/\hbar$.

\subsection{Electromagnetic waves}
The wave equations 
\begin{equation}
\nabla_\perp^2 {\bf E}_{\rm sf} = \mu_v\epsilon_v\frac{\partial^2 {\bf E}_{\rm sf}}{\partial t^2} 
\end{equation}
and
\begin{equation}
\nabla_\perp^2 {\bf B}_{\rm sf} = \mu_v\epsilon_v\frac{\partial^2 {\bf B}_{\rm sf}}{\partial t^2} 
\end{equation}
may be derived directly using the hydrodynamic Maxwell equations, Eq.~(\ref{GPgausse}), Eq.~(\ref{GPgaussb}), Eq.~(\ref{GPfara}) and Eq.~(\ref{GPmax}) in the usual way. In the long wave length limit the waves described by these equations are associated with the linearised (infinitesimal) density perturbations of the Gross--Pitaevskii equation \cite{pethick2002bose,pita2003bose}
\begin{equation}
\nabla^2 \delta n({\bf r}) = \frac{1}{c^2_s}\frac{\partial^2 \delta n({\bf r})}{\partial t^2}.
\end{equation}
These Bogoliubov phonons have the well known dispersion relation
\begin{equation}
\omega_{\rm phonon} = \sqrt{\left(\frac{p^2}{2m} \right)^2 +\frac{c_0n}{m} p^2},
\end{equation}
which, in the long wave length limit, is a linear function $E=pc_s$ of momentum $p = \hbar k$ with the constant of proportionality $c_s=\sqrt{c_0n/m}$ equal to the speed of sound. These sound waves have the remarkable property that they also correspond to a propagating density perturbation of the condensate and thereby a spatiotemporal variation in the quantum pressure, which will later be associated with gravity.

\subsection{Quantum electrodynamics}

The preceeding assumptions meant that vortex charges were frozen in the superfluid and this lead to the Maxwellian approximation of classical electrodynamics. However, accounting for the fact that the superfluid is compressible intrinsically enables vortex-antivortex pair creation and annihilation events that are accompanied by emission and absorption of phonon radiation \cite{Parkersound,Groszek2016a}. Indeed, the full theory, Eqns.~(\ref{GPEegy}) and (\ref{GPEtd}), self-consistently describe the processes of relativistic quantum electrodynamics \cite{Popov1973a,Klein2014a} as can readily be observed in numerical simulations, shown e.g. in the supplementary movie of Ref.~\cite{Orderfromturbulence}. 

It is also possible to go beyond quantum electrodynamics for example by introducing a spin degree of freedom to the superfluid, which enables the creation of new kinds of vortex particles that may behave as non-Abelian anyons \cite{PhysRevLett.123.140404}. The resulting superfluid quantum chromodynamics has connections to the field of topological quantum computation \cite{TQC}. We will next move on to consider the $\Psi^2_{\rm G}$ term and the emergence of gravity. 

\section{Emergent Gravity}

The Einstein field equations of general relativity can be derived using a Lagrangian variational principle with the matter free part of the four dimensional spacetime generated by the Einstein--Hilbert action \cite{misner1973gravitation}
\begin{equation}
S_{\rm EH} = \int \mathcal{L} d{\bf r}^4= \frac{c^4}{16\pi G}\int R\sqrt{-g} d{\bf r}^4,
\label{EHaction}
\end{equation} 
where $R$ is the Ricci scalar and $c$ is the speed of light, $G$ is Newton's gravitational constant and the integration is over four-dimensional spacetime coordinates. Sakharov took the viewpoint that the Lagrangian $\mathcal L$ would be generated by an underlying quantum field theory and expressed the vacuum quantum fluctuations as a series expansion
\begin{equation}
\mathcal{L} = \lambda + \alpha R  + \beta R^2\ldots
\label{Lagrangian}
\end{equation} 
where the first term in the right corresponds to the cosmological constant, the second term gives rise to the Einstein--Hilbert action that yields the Einstein's field equations and the remaining terms result in higher order corrections to general relativity \cite{Sakharov1967a,misner1973gravitation}. In this picture, gravity and general relativity are emergent phenomena generated by vacuum fluctuations (or quantum turbulence) of the underlying quantum field theory. Gravity in the superfluid universe and the gravitational mass of quantised vortices have similar origin.

The linear Bogoliubov phonon dispersion relation may be described in terms of an acoustic metric \cite{Unruh1981a,Visser2005a} of the superfluid universe
\begin{equation}
g_{\mu\nu} = {\Omega}^2
\left(
\begin{matrix}
-(c_s^2 -v_s^2) &\vline& -v_j \\
\hline
-v_i &\vline& \delta_{ij}
\end{matrix}
\right)
\end{equation}
where the conformal factor $\Omega$ is constant for flat spacetime. For the curved spacetime with two space dimensions, relevant to our discussion, $\Omega = mn/c_s$ and 
\begin{equation}
\sqrt{-g} = \Omega^2c_s=\frac{m^2n^2}{c_s},
\end{equation}
where $g=\det({g_{\mu\nu}})$ denotes the determinant of the metric tensor. The spacetime interval \cite{Unruh1981a,Visser2005a}
\begin{equation}
ds^2 = \Omega^2[-c_sdt^2 + (d{\bf x} - {\boldsymbol v}_sdt)^2]
\end{equation}
accounts for the linear part of the Bogoliubov phonon dispersion relation.

Gravity in the emergent matter free superfluid universe arises due to the quantum fluctuations (superfluid wave turbulence) that results in the condensate density fluctuations even at zero temperature due to the fluctuating quantum depletion, caused by the trans-Planckian (atom-atom) particle interactions \cite{Lee1957a,pethick2002bose,pita2003bose}. We begin by expressing the gravitational energy
\begin{equation}
{\rm GEM}_{\rm G}  =\int \frac{\hbar^2}{2m}\left(\nabla\left|\Phi({\bf r})\right| \right)^2   d{\bf r}^2,
\label{gemg}
\end{equation}
in terms of the quantum pressure
\begin{equation}
P_q = -\frac{\hbar^2}{2m\sqrt{n}} \nabla^2\sqrt{n} = -\frac{\hbar^2}{4m} \nabla^2 \ln\left(\frac{n}{n_0}\right) -\frac{\hbar^2}{2m}\left( \frac{|\nabla \sqrt{n}|}{\sqrt{n}}\right)^2.
\label{QP}
\end{equation}
This yields
\begin{equation}
{\rm GEM}_{\rm G}  = -\int n\left[\frac{\hbar^2}{4m} \nabla^2 \ln\left(\frac{n}{n_0}\right)+P_q \right] d{\bf r}^2. 
\label{gemgqp}
\end{equation}
In two-dimensional space the Riemann tensor reduces to the Ricci scalar, which is related by $K=R/2$ to the geometric Gaussian curvature $K$. Combining Liouville's equation of differential geometry
\begin{equation}
 \nabla^2\ln(\tilde{\Omega}) = -K\tilde{\Omega}^2,
\end{equation}
where $\tilde{\Omega}=\Omega/\Omega_0$, with Eq.(\ref{gemgqp}) we obtain
\begin{equation}
{\rm GEM}_{\rm G}  = \int n\left[  \frac{\hbar^2}{8m} R \tilde{\Omega}^2 -P_q  \right] \ d{\bf r}^2, 
\label{GEM4}
\end{equation}
which may also be expressed as
\begin{equation}
{\rm GEM}_{\rm G}  =  \frac{\hbar^2 c_s}{8 m^3n_0^2}\int\sqrt{-g}\left[ nR  +\frac{4\sqrt{n}}{\tilde{\Omega}^2} \nabla^2\sqrt{n}  \right] \ d{\bf r}^2. 
\label{GEM5}
\end{equation}
Associating the Lagrangian of the two-dimensional superfluid universe with the quantum kinetic energy 
$
{\rm GEM}_{\rm G}=\int\mathcal{L}_{\rm G}^{(2+1)}d{\bf r}^2  
$
brings about the connection to the superfluid Einstein--Hilbert action
\begin{align}
S_{\rm SEH}& = \int \mathcal{L}_{\rm G}^{(2+1)} d{\bf r}^2c_sdt \notag\\
&= \frac{c_s^4}{16\pi }\int\sqrt{-g}\left[\phi R  -\frac{\omega}{\phi}\frac{ \nabla^2\sqrt{\phi} }{\sqrt{\phi}} \right] \ d{\bf r}^2c_sdt,
\label{EHactionsf}
\end{align} 
where 
\begin{equation}
\phi=\frac{8\pi\hbar^2 n}{m^3c_s^3n_0^2} \;\;\;{\rm and}\;\;\; \omega = -\frac{32\pi\hbar^2}{m^3c_s^3}. 
\end{equation}
The structure of this action where curvature is coupled to a (dark matter) scalar field bears similarity to the Jordan--Brans--Dicke dilaton theories \cite{Brans2005a,GRUMILLER2002327}, which are a broader class of scalar-tensor theories of gravity considered already by Gunnar Nordstr\"om \cite{Gunnar1913a,Ravndal2004a}. Here the scalar field $\phi$ couples to all matter and energy fields and infact is the ``source of everything" in the superfluid universe. Indeed, it is straightforward to add all of the matter and energy fields $\mathcal{L}_{\rm EM,DM,NM,DE}$ in Eq.~(\ref{GP22}) to the `vacuum' action (\ref{EHactionsf}), including the effects of the dark energy/cosmological constant via $\mu_{\rm DE}$. The superfluid gravitational field is non-zero in any region of space where the condensate particle density is spatially varying. As such, the density fluctuations inherent to two-dimensional quantum turbulence may be interpreted as corresponding to vacuum fluctuations of the superfluid universe. 

In a matter free universe quantum fluctuations yield spacetime curvature (via modulation of $n$) locally. Gravitation in the large-scale structure of the universe can be `added' by introducing an `external' potential. For example, a harmonic trapping potential would yield a non-uniform Thomas--Fermi condensate density, where as a self-gravitating universe could be induced by long-ranged dipole-dipole interactions that have been observed to generate self-bound droplets \cite{Pfau2016a}. Antitrapping external potentials (cosmologies) are also frequently used in cold atom experiments. Generically, it is possible to imprint any density landscapes in the laboratory condensates, such as a ``Bose--Einstein cosmology" \cite{Gauthier2016a}. 

\subsection{Gravitational waves}

In the context of general gravity in 2+1 dimensions the `folklore' states that due to the lack of degrees of freedom, the theory should be trivial and that there should be no gravitational waves, and that gravity would then be manifest only via topological effects \cite{WITTEN198846,carlip1993ways}. However, in Eq.~(\ref{EHactionsf}) compressibility of the dark matter field provides the local degrees of freedom absent in the Einsteinian 2+1 dimensional gravity. It is therefore reasonable to anticipate the possibility of wave motion akin to gravitational waves to exits in this superfluid universe similar to the higher dimensional Jordan--Brans--Dicke theories. The question is then what should such waves physically correspond to in laboratory experiments? The ``desirable" properties of such waves might be that their speed of propagation be close to the speed of sound $c_s$ and that the generalised angular momentum they carry would be two. For a ground state scalar BEC there exists only one gapless excitation branch linear in momentum---the usual Bogoliubov phonon. However, those quasiparticles, although producing spatial modulations of the condensate density, are plane waves carrying an angular momentum of zero and therefore the otherwise plausible idea of associating the longest wavelength phonons as gravitational waves in this system seems tenuous. Another idea would be to associate other excitations, such as the scissors modes, pertinent to the quadrupole operator with the gravitons because these have angular momentum quantum number two but this would trade off the gapless linear spectrum. The next possible direction could be to consider gravitational waves as a genuinely non-linear effect and to associate them with two-dimensional Jones--Roberts solitons, or other vortexonium-like rarefaction pulses, that have a dispersion relation at low momenta whose slope does coincide with that of the phonons \cite{1982JPhA...15.2599J,Groszek2016a}. In the presence of matter (e.g. vortex lattices) yet another possibility arises. The Kelvin--Tkachenko vortex shear waves are new excitation modes below the phonon line, also linear in momentum in the ``stiff" limit \cite{Coddington2003a,Baym2003a,Simula2013a}. These are transverse shear waves that correspond to the collective motion of the vortex particles and may be viewed as the mean-field precursors for the collective degrees of freedom that yield a geometric description of the fractional quantum-Hall effect \cite{Haldane2011a,Haldane2019a}. However, we shall deem more rigorous contemplations of the nature of gravitons in the superfluid universe model to be outside the scope of this study.

\subsection{Gravity, topology and enstrophy}

In the superfluid universe, topology of the spacetime is inherently linked to the spacetime curvature. At zero temperature the condensate groundstate is smooth and the universe is composed of dark matter only. If quantum turbulence is triggered, e.g. via a parameter quench or tunneling to a lower energy state, particles (vortices), electromagnetic field (condensate phase gradients) and gravity (condensate density gradients) all emerge. The topology of such a multiply connected condensate and the resulting  gravity are linked by the Gauss--Bonnet theorem
\begin{equation}
\int_\mathcal{M} KdS + \int_{\partial\mathcal{M}} k_g dl= 2\pi\zeta(\mathcal{M}), 
\label{GB}
\end{equation}
which  is a statement that the sum of the total curvature of a compact 2D Riemannian manifold $\mathcal{M}$, and the rotation of its smooth surface ${\partial\mathcal{M}}$ is proportional to the Euler characteristic $\zeta$ of $\mathcal{M}$. For a planar BEC with equal number of vortices and antivortices, Eq.~(\ref{GB}) reduces to
\begin{equation}
\int_\mathcal{M} KdS = 2\pi(1-g_t),
\label{GB2}
\end{equation}
where $g_t$ is the genus of the surface. For a ground state condensate $g_t=0$ and generically $g_t=N_v$ for a quantum turbulent BEC with $N_v$ vortices. In general, due to the quantization of circulation, the number of vortices is also related to the enstrophy of the system of single quantum vortices
\begin{equation}
\mathbb{E}=\int |\nabla\times {\boldsymbol v}_s |^2 dS=\kappa^2 g_tf,
\end{equation} 
where $N_v$ is the number of single quantum vortices in the system and the generalised function $f=\int\delta({\bf r} -{\bf r}')\delta({\bf r} -{\bf r}')dS$. To be consistent with the Feynman rule, the coarse grained average $\langle f\rangle = 4n_v$.
The enstrophy, a purely topological entity here, is thus a measure of the total curvature 
\begin{equation}
\mathbb{E}= \kappa^2(1- \frac{1}{4\pi}\int RdS)f,
\label{ens}
\end{equation}
linking an important hydrodynamical quantity to gravity. In the theory of two-dimensional turbulence, the conservation law of enstrophy, $\frac{\partial \mathbb{E}}{\partial t}=0$, underpins the inverse energy cascade, which ultimately leads to the phase separation of vortices and antivortices into Onsager vortex clusters, and a seeming matter-antimatter asymmetry in the theory \cite{EmergePRL}. 

Equation~(\ref{ens}) shows that regions of high enstrophy, such as occurs within Onsager vortices \cite{Gauthier2019a,Johnstone2019a}, may also correspond to regions of high curvature. This naturally leads to the interpretation that the Einstein--Bose condensation transition \cite{Orderfromturbulence,Valani2018a} at negative absolute temperature would be expected to lead to the formation of an analogue black hole with the associated phenomenology such as event horizons, ergo regions, Hawking radiation and black hole thermodynamics. 

\section{Gravitational vortex mass}
Equipped with the preceeding considerations we are in a position to discuss the gravitational mass of a vortex. Adding a quantised vortex in an otherwise flat superfluid universe, $n(r)=n_0={\rm const}$, changes the topology of the spacetime and influences its dynamics. The qualitative new features brought about by the nucleation of a vortex include: 
\begin{itemize}
\item[(i)] the topology of the condensate changes from being singly connected to being multiply connected,
\item[(ii)] a new vortex core bound quasiparticle---kelvon that is a component of the normal matter of the superfluid universe---emerges in the elementary excitation spectrum,
\item[(iii)] the vortex acquires a mass due to its coupling to the dark matter field,
\item[(vi)] the superfluid vacuum begins to flow due to the phase gradient of the condensate and this superflow corresponds to an emergent electric field  
\item[(vii)] motion of the vortex (kelvon) induces a magnetic field due to the time variation of the condensate density, 
\item[(viii)] accelerating vortex may radiate phonons, and
\item[(ix)] a condensate density gradient due to the structure of the vortex core results in quantum pressure that gives rise to a gravitational field. \end{itemize}

A vortex centered at the origin may be described by the wavefunction
\begin{equation}
\psi_v(r) = n_0\chi(r) e^{iS(r,t)}
\end{equation}
where $\chi$ is the vortex core structure function \cite{Fetter1971a, PhysRevX.2.041001}
\begin{equation}
\chi(r) = \tanh\left(\frac{r}{\sqrt{2}r_c}\right) \approx \frac{r}{\sqrt{r^2+r_c^2}}
\end{equation}
and the phase function $S = \arctan(x,y)$ has a singularity at the origin. In the vicinity of the vortex core the condensate density $n(r)$ forms a harmonic oscillator potential 
\begin{equation}
n(r)_{r\to 0}=n_0\chi^2_{r\to 0}=n_0 r^2/r_c^2.
\end{equation}
A test vortex with circulation $q_2$ placed distance $r$ from the origin is influenced by two forces due to the presence of the source vortex of circulation $q_1$ at the origin:
(i) the electric force ${\bf F}_{e}$ due to the phase gradient (superflow) and (ii) the gravitational (Magnus) force ${\bf F}_{g}$ due to the local density gradient (quantum pressure). The forces ${\bf F}_{e}$ and ${\bf F}_{g}$ may be obtained by considering the difference in energy, $\Delta{\rm GEM}=(\mathcal{E}_1- \mathcal{E}_0)$, between a universe with and without a vortex \cite{pethick2002bose}
\begin{equation}
\Delta{\rm GEM} =  \int \frac{\hbar^2n_0}{2m} \left[\left(\frac{\chi}{r}\right)^2   +\left(\frac{\partial \chi}{\partial r}\right)^2  + \frac{1}{2}(1-\chi^2)^2   \right]d{\bf r}^2.
\label{GPDa}
\end{equation}
We associate the first two terms in Eq.~(\ref{GPDa}), respectively, with the electric and gravitational fields
produced by the vortex particle, while the last term is due to the change in the dark matter energy density. It would be tempting to elevate these terms into potentials the negative gradients of which would then yield the conservative forces on the vortex. In the case of the electric force, the integral of the first integrand in Eq.~(\ref{GPDa}) does indeed yield the usual  electrostatic potential proportional to $\ln(r/r_c)$ in two-dimensions. However, for the second, gravitational, term this approach only works for the short distance limit as clarified further below. 

In contrast to classical electrodynamics where the Lorentz force is determined solely by the electromagnetic fields acting on the charged particle and is independent of all other forces, Eq.~(\ref{Lorentz}) completely determines the dynamics of a vortex. The Lorentz force 
\begin{align}
{\bf F}_{\rm Lorentz}={\bf F}_{\rm Magnus}=-{\bf F}_{\rm g}
\label{LorMag2}
\end{align} 
is identified as the Magnus force of superfluid hydrodynamics but more interestingly, it is also the negative of the gravitational force. This means that it is not possible to independently vary the gravitational and electromagnetic forces acting on a vortex since changing phase gradients induce density gradients and vice versa. Formally this is expressed by the relation
\begin{equation}
{\bf F}_{\rm g}+{\bf F}_{e} + {\bf F}_{m}=0,
\end{equation}
where ${\bf F}_{m}$ is the magnetic component of the electromagnetic force, the last term in Eq.~(\ref{Lorentz}).
In words, gravity, electricity and magnetism are the `three sides of the same coin'.

A strictly uniform system, ${\boldsymbol v}_n=0$, corresponds to a zero gravity ${\bf F}_{g}=0$ because the Magnus force vanishes and the vortex is frozen in the superfluid traveling at the speed of local superfluid velocity such that ${\boldsymbol v}_v={\boldsymbol v}_S$. Consequently, ${\bf F}_{m}= -{\bf F}_{e}=m^v_i a$, where $a$ is the inertial acceleration of the vortex. For non-uniform systems, such as harmonically trapped condensates, gravitational effects become important and the Magnus force has a non-zero value. For an infinite system with a vortex placed in a region of a parabolic underdensity ${\boldsymbol v}_S=0$ due to the image of the vortex being infinitely far from the vortex such that ${\boldsymbol v}_v={\boldsymbol v}_n$. In this case ${\bf F}_{e}=0$ and ${\bf F}_{m}=-{\bf F}_{g}=m^v_g g$, where $g$ is the gravitational acceleration of the vortex. The resulting periodic circular vortex motion is then entirely due to the curvature of the condensate density. We may then consider placing the vortex in ``Einstein's elevator" such that it is not possible to distinguish between the two aforementioned cases, which may be set up such that $a=g$. Hence we arrive at a vortex equivalence principle---the equality of the inertial and gravitational vortex masses:
\begin{equation}
m^v_{\rm g }=  m^v_{\rm i }.
\label{gravinert}
\end{equation}
This may not be surprising since both gravitation and electromagnetism in this theory are generated by the same emergent quantity---the Laplacian of the matterwave of the Bose--Einstein condensate. Furthermore, the mass $m^v$ and charge $q
_v$ of the vortex are not independent quantities but are related by
\begin{equation}
\frac{q_v}{m^v}=\epsilon_v \omega_k  = \frac{h}{m_am_v}
\end{equation}
where $m_a=m$ is the mass of the atom. The kelvon frequency thus combines the masses of the dark matter particles that form the superfluid and the elementary quasiparticles of the superfluid universe.

The explicit forms of the forces on a test vortex are obtained directly from the respective terms in the Lorentz force Eq.~(\ref{Lorentz})
\begin{align}
{\bf F}_{e} &=-mn_{bg} 
{\boldsymbol \kappa} \times{\boldsymbol v}_S\notag\\ &=\frac{2\pi\hbar^2n_0\chi^2}{m}|\nabla_\perp S| {\bf e}_{12}= \frac{q_{1}q_{2}}{2\pi\epsilon_v}\frac{r}{r^2+r^2_c}{\bf e}_{12},
\end{align}
where ${\bf e}_{12}$ is a unit vector from the source vortex to the test vortex, $q_i=\pm h/m$, and the gravitational force
\begin{align}
{\bf F}_{g} &=-mn_{bg} {\boldsymbol \kappa} \times{\boldsymbol v}_n \notag \\&= -\frac{\pi\hbar^2n_0\chi^2}{m}\frac{\partial_r (\chi^2)}{\chi^2}{\bf e}_{12}=-G_vm_1m_2\frac{r r_c^2}{(r^2+r^2_c)^2}{\bf e}_{12}.
\end{align}
As in Einsteinian gravity, here too we have judiciously defined the gravitational constant, $G_v=\omega_k^2/4\pi mn_0$, in such way that the gravitational vortex mass is, by construction, equal to the inertial vortex mass as stipulated by the superfluid equivalence principle.

The electric force is repulsive if the test vortex has the same sign of circulation as the source vortex, and attractive if the test vortex has an opposite sign of  circulation with respect to the source vortex. All vortices have a negative mass \cite{Yefsah2013a,Simula2018a} such that the product $m_{1}m_{2}$ is always positive and the vortex-vortex gravity is always an attractive interaction. This is because all vortices create a parabolic underdensity in the condensate. However, considering for instance a bright soliton and a vortex would result in a repulsive gravitational interaction between the two. For short distances, $r<r_c$ the forces reduce to
\begin{align}
{\bf F}_{e} = \frac{q_{1}q_{2}}{2\pi\epsilon_v}\frac{r}{r^2_c}{\bf e}_{12},
\label{fe}
\end{align}
and
\begin{align}
{\bf F}_{g} &=-G_vm_1m_2\frac{r}{r^2_c}{\bf e}_{12}.
\label{fg}
\end{align}
These approximate ``Newtonian" forces, Eqns.~(\ref{fe}) and (\ref{fg}) can be obtained, in the $r<r_c$ limit, as the negative gradients of the respective potentials both being $\propto r^2$ (the first two terms of Eq.~(\ref{GPDa})). However, for large distances the $1/r^3$ behaviour of gravity is very different from the $1/r$ force law of the electromagnetic field.

\section{Vortex--Particle duality}

Dualities in physics are a powerful concept that grant multiple viewpoints for the same physical phenomenon. Prominent examples of such dualities include the anti-deSitter/conformal field theory (AdS/CFT) correspondence and its variants, holographic principle, gauge-gravity duality, bulk-edge correspondence, and fluid-gravity duality \cite{Maldacena1998,ammon_erdmenger_2015}. 

The two-dimensional superfluid universe features a vortex--particle duality. The vortex--particle duality enables a description of the same system either in terms of the two dimensional condensate atoms comprising the superfluid with vortices embedded in the real space of atoms or in terms of the one dimensional vortex particles with condensate atoms embedded in the phase space  of the vortices \cite{Simula2019book}. Each of these two sides of the duality are characterised separately below.  

\subsection{Two-dimensional weakly interacting classical field theory with gravity}
The dynamics of the superfluid (atoms) moving in two-dimensional space is described by the Gross--Pitaevskii equation, Eq.~(\ref{GPEtd}). Each of the atoms have four ($q_x,q_y,p_x,p_y$) canonical phase space coordinates such that the Hamilton's equations of motion are
\begin{align}
 \dot{q}_x &= -\frac{\partial H_{\rm atom}}{\partial p_x} 
   \hspace{1cm}
    \dot{q}_y = -\frac{\partial H_{\rm atom}}{\partial p_y}\\\nonumber
    \dot{p}_x &= \frac{\partial H_{\rm atom}}{\partial q_x} 
    \hspace{1cm}
    \dot{p}_y = \frac{\partial H_{\rm atom}}{\partial q_y},
\end{align}
where $H_{\rm atom}$ is a Hamiltonian for the atoms.

When quantised vortices are nucleated in the superfluid, the condensate order parameter becomes topologically multiply connected. The vortices that puncture the condensate are not part of the fluid although their motion is fully correlated with that of the fluid. The interaction between the fluid and the vortices is mediated by the dark matter. The atoms ``experience" the vortices as obstructions that constrain their dynamics and the vortices ``experience" the fluid of atoms as an obstruction that they have to plough through. The vortices acquire mass due to their interaction with the Higgs-like dark matter field (the condensate).

The description of the two-dimensional fluid with its four dimensional phase space corresponds to a (2+1)-dimensional weakly gravitating classical field theory for which the vortex degrees of freedom realize a (1+1)-dimensional boundary quantum field theory. The gravity that originates from the quantum pressure of the condensate is emergent and is produced by e.g. quantum turbulence. The importance of the fact that the fluid atoms and the vortex particles (kelvons) formally exist in spacetimes of different dimensionality cannot be overemphasised.

\subsection{One-dimensional strongly interacting quantum field theory without gravity}
The dynamics of the vortices is described by the vortex equation of motion Eq.~(\ref{vortexeqn}) to which the Onsager point vortex model provides a rather good approximation in the dilute vortex gas limit. Each of the vortices have two ($q_x,p_x$) canonical phase space coordinates such that the Hamilton's equations of motion are
\begin{align}
 \dot{q}_x = -\frac{\partial H_{\rm vortex}}{\partial p_x} 
     \hspace{1cm}
    \dot{p}_x = \frac{\partial H_{\rm vortex}}{\partial q_x}. 
\end{align}
The function $H_{\rm vortex}$ is the well known two-dimensional Coulomb gas pseudo Hamiltonian
\begin{equation}
H_{\rm vortex} = -\sum_{i<j}^{N_v}s_is_j\ln(|{\bf r}_i-{\bf r}_j|/r_c),
\end{equation}
where ${\bf r}_i = \sqrt{{\bf q}_{x,i}^2+{\bf p}_{x,i}^2}$ and which can be further mapped either onto the two-dimensional classical XY model (fluid picture) or to the one-dimensional sine-Gordon quantum field theory (particle picture) \cite{giamarchi2004quantum}. Quantum mechanically, the vortex particles correspond to the quantised kelvon quasiparticles entering the Eq.~(\ref{rho}). In the presence of a circular boundary and in the vicinity of the Einstein--Bose condensation transition, the point-vortex model can be mapped onto an inverted, strongly interacting one dimensional harmonic oscillator Hamiltonian \cite{Valani2018a}.

\section{Quantum Hall effects}

A Hamiltonian for an electron in a uniform magnetic field ${\bf B} =\nabla \times {\bf A}_e$ is
\begin{equation}
    H_e = \frac{({\bf p }-q_e{\bf A}_e)^2}{2m_e},
    \label{eham}
\end{equation}
where $m_e$ is the mass of the electron, $q_e$ its charge, ${\bf p}$ its momentum and ${\bf A}_e$ is the vector potential. When the magnetic field strength $B$ is sufficiently increased, a two-dimensional electron gas with fixed number of electrons undergoes successive topological quantum phase transitions to strongly correlated integer and fractional (when Coulomb interactions are accounted for) quantum Hall liquids \cite{Thoulessbook}. Such topological states of matter are anticipated to emerge when the filling fraction
\begin{equation}
\nu_e = \frac{N_e}{\Phi/\Phi_0}=\frac{N_e}{N_\Phi} \lesssim 1,
\label{nue}
\end{equation}
where $N_e$ and $N_\Phi$ are the number of electrons and the number of magnetic flux quanta, respectively, and $\Phi = B\mathcal{A}$ is the magnetic flux piercing area $\mathcal{A}$ and quantised in units of $\Phi_0=h/2e$.

A great effort has been dispensed in trying to observe bosonic quantum Hall states using rapidly rotating neutral superfluids \cite{pethick2002bose,WilkinGunn,Cooper2008a,FetterRevModPhys,Schweikhard,SpielmanNature2009}. This has been prompted by the observation that the Hamiltonian of such systems can be mapped onto that of the two-dimensional electron problem, Eq.~(\ref{eham}). Specifically, a Bose--Einstein condensate in a harmonic oscillator potential, expressed in the rotating frame of reference, has the `single-particle' Hamiltonian
\begin{align}
    H_a &= \frac{{\bf p}^2 }{2m_a} +\frac{1}{2}m\omega_{\rm osc}^2 r^2 +gn -\Omega_{\rm rot} L_z \nonumber \\
    &=\frac{({\bf p }-q_a{\bf A}_a)^2}{2m_a}+\frac{1}{2}m[\omega_{\rm osc}^2-\Omega_{\rm rot}^2] r^2 +gn,
    \label{becham}
\end{align}
where $m_a$ is the mass of the atom, $\omega_{\rm osc}$ is the harmonic oscillator frequency, $\Omega_{\rm rot}$ is the external rotation frequency and $L_z$ is the axial component of the orbital angular momentum operator. When $\Omega_{\rm rot}$ is increased and is approaching the value of $\omega_{\rm osc}$ such that $\omega_{\rm osc}^2-\Omega_{\rm rot}^2\to0$, ever larger number of vortices are nucleated in the system while the atom cloud expands radially becoming ever more dilute such that $n\to0$. The result is that the last two terms in Eq.~(\ref{becham}) become negligible with respect to the first term such that $H_a$ becomes mathematically identical to $H_e$ with
\begin{equation}
q_a{\bf A}_a = -m_a\Omega_{\rm rot} {\bf r}\times {\bf e}_z.
\label{qA}
\end{equation}
Based on this observation, it is often stated that the rotation frequency $\Omega_{\rm rot}$ of a neutral superfluid would correspond to the magnetic field $B$ in the problem of a two-dimensional electron gas. Consequently, to realize quantum Hall states, the goal would be to try to make $\Omega_{\rm rot}$ as large as possible in order to reach the limit of strong effective magnetic fields, analogously to the case of degenerate two-dimensional electron gas in a strong external magnetic field. However, we argue that increasing $\Omega_{\rm rot}$ actually diminishes the effective magnetic field and increases the filling factor, which in correspondence with Eq.~(\ref{nue}) should be 
\begin{equation}
\nu_v = \frac{N_v}{N_a},
\end{equation}
as opposed to its inverse \cite{Cooper2008a,FetterRevModPhys}, where $N_a$ is the number of atoms and the flux quantum corresponds to the mass of an atom $m_a$. Zeroing the effective trapping potential and particle interactions in Eq.~(\ref{becham}) can clearly be achieved by setting $\Omega_{\rm rot}=\omega_{\rm osc}$ but this is not the same as increasing ${\bf A}_e$ alone in Eq.~(\ref{eham}). The analogue of electric charge should be the quantum of circulation \cite{Simula2012a}, such that for a rotating BEC the effective total charge 
\begin{equation}
q_a=\kappa N_v.
\end{equation}
Substituting this and the Feynman rule, Eq.~(\ref{feyn}), to Eq.~(\ref{qA}), we obtain the vector potential per condensate particle
\begin{equation}
{\bf A}_a={\bf A}_v/N_a = -\frac{m_a}{2\mathcal{A}} {\bf r}\times {\bf e}_z,
\label{A}
\end{equation}
such that the magnetic field is
\begin{equation}
{\bf B}_v =\nabla\times {\bf A}_v = \frac{m_aN_a}{\mathcal{A}} {\bf e}_z,
\label{B}
\end{equation}
and has no explicit dependence on $\Omega_{\rm rot}$. Similarly, the charge $q_a$ is only implicitly dependent on $\Omega_{\rm rot}$ and it is only in the product of $q_a{\bf A}_a$ that the rotation frequency makes an explicit appearance. We also note that ${\bf B}_v=m_an{\bf e}_z$ as expressed in Eq.~(\ref{B}) follows directly from the definition of ${\bf B}_{\rm sf}$ in Eq.~(\ref{EB}), see also below Eq.~(\ref{Lorentz}).

When $\Omega_{\rm rot}$ increases, $q_aA_a$ increases because $q_a$ grows, even though both $A_a$ and $B_a$ \emph{decrease} due to the increasing area $\mathcal{A}$ occupied by the superfluid. Nevertheless, $N_v$ grows faster than $\mathcal{A}$ because also the vortex density increases. Hence, the filling factor criterion, Eq.~(\ref{nue}), to achieve quantum Hall limit can be expressed as
\begin{equation}
\nu_v = \frac{N_v}{B_a \mathcal{A}/m_a}=\frac{N_v}{N_a}\lesssim 1,
\end{equation}
which in practice means that this criterion is immediately satisfied when the first vortex is nucleated in the system. However, for low rotation frequencies the lowest Landau level states are not degenerate and thus do not form a flat band because of the non-negligible influence of the last two terms in Eq.~(\ref{becham}) that lift the degeneracy, and it is for this reason that the system needs to be rotated rapidly to make the kinetic energy overwhelm the scalar potentials---instead of creating a large number of vortices per se. In fact, as previously mentioned, when the system is rotated faster, the vortex number goes up and this causes the value of $\nu_v$ to \emph{increase}.

This also means that the rotation frequency $\Omega_{\rm rot}$ should not be associated with a magnetic field, rather, the rotating drive creates a strong electric field, Eq.~(\ref{EB}), (superflow), which destabilizes the vacuum and nucleates an increasing number of charges (vortices) in the system in accordance with Eq.~(\ref{feyn}). In contrast, the number of electrons does not change when a two-dimensional electron gas is placed in a magnetic field in typical quantum Hall effect experiments.

Associating vortices with charges and magnetic flux with atoms (rather than vice versa) is further supported by the prediction that a vortex transported around a loop $\mathcal{C}$ in a superfluid accumulates a Berry phase \cite{HaldaneBerry,AoThoulessBerry}
\begin{equation}
\gamma_\mathcal{C} = \frac{2\pi}{m_a} \int_{\mathcal{A}_\mathcal{C}} {\bf B}_v\cdot d{\mathcal A} = 2\pi N_a(\mathcal{C}),
\end{equation}
where $N_a$ is the number of atoms enclosed by the vortex path. As such, the atoms are the Aharonov--Bohm flux quanta for the vortex. The force on a vortex executing circular motion in a perpendicular magnetic field is
$
   {F}= \kappa{ v}_v { B}_{\rm sf}=m_vv^2_v/r
$
such that the vortex mass can be expressed in terms of the geometric phase as
\begin{equation}
    m_v = \frac{\gamma_\mathcal{C}}{\mathcal{A}_\mathcal{C}}\frac{\hbar}{\omega_v}.
    \label{massberry}
\end{equation}
Although the possiblity of fractional statistics for a single vortex was excluded in Ref.~\cite{HaldaneBerry} it may be possible that many vortex Kelvin--Tkachenko states would allow it as the vortices then form collective composite quasiparticles in analogue with the flux attachment to many-electron states in the fractional quantum Hall effect (FQHE). 

The above reasoning leads to a proposition for a potentially new interpretation of the Bogoliubov quasiparticle excitation spectrum of rotating condensates. Since in the quantum-Hall limit the number of lowest Landau level (LLL) eigenstates should correspond to the number of flux quanta, we are motivated to re-define the vortex filling factor by associating the number of populated LLL states with the magnetic flux;
\begin{equation}
\nu_{\rm LLL} = \frac{N_v}{N_{\rm LLL}}
\end{equation}
so that the FQHE states with $\nu_{\rm LLL}$ should be associated with condensates of elementary droplets consisting of $N_v$ vortices bound to $N_{\rm LLL}$ one-particle Landau level states \cite{Haldane2011a,girvin_yang_2019}. Thus we arrive at the following interpretation of the physics of a rotating BEC: in addition to the phonons, the low-lying quasiparticle excitation spectrum comprises two types of modes (i) the  surface modes which approximately correspond to single particle harmonic oscillator angular momentum eigenstates that ultimately as $\Omega_{\rm rot}\to\omega_{\rm osc}$ will form the LLL and (ii) $N_v$ transverse vortex shear modes (Kelvin--Tkachenko modes) visualised e.g. in the supplementary videos of Ref.~\cite{Simula2013a}. Each of these vortex shear modes can be reconstructed as a product of $N_{\rm LLL}$, non-degenerate, LLL states and as such should be viewed as the meanfield precursors to the 
many-body FQHE states that could be realised experimentally as metastable excited states of rotating BECs. We aim to return to this point in more detail elsewhere merely emphasising again in this context that rapidly rotating a neutral superfluid should ultimately lead to: 
\begin{itemize}
    \item[(i)] the formation of Onsager vortex clusters and the associated absolute negative temperature states,
    \item[(ii)] the formation of fractional quantum-Hall-like states due to the quasiparticle condensation in the hierarchy of transverse vortex wave modes, and
    \item[(iii)] the formation of an analogue black hole due to the increasing mass density of vortex particles. 
\end{itemize}
This means that the dense vortex matter may be described in terms of at least three complementary pictures, the hydrodynamical, the electromagnetic, and the gravitational. Indeed, in light of identifying the quantised vortices as charged massive particles, the connections between the physics of black holes and FQHE \cite{Haldane2011a,Wiegmann,Hedge2019a} are perhaps less surprising.

\section{Conclusions}

We have considered an emergent (2+1) dimensional superfluid universe where gravity and electromagnetism have the same origin, the quantum kinetic energy of the superfluid, and are coupled to the dark matter field, which represents the fabric of the superfluid spacetime. In this universe, electric field corresponds to the superflow of the Bose--Einstein condensate, magnetic field corresponds to the condensate phase evolution, vortices are massive charged particles and the sound waves correspond to the massless photons. Gravity is associated with condensate density gradients and the condensate is identified as the elusive dark matter. 

The vortices have two possible signs of circulation and therefore the electromagnetic interaction between two vortices may be attractive or repulsive. Both vortices and antivortices have the same condensate density depletion in their cores and, correspondingly, the density gradients produced by them are identical. Therefore, the gravitational interaction between any two vortices is always attractive. The vortex acquires a mass by interacting with a Higgs-like dark matter field whose density, together with the fundamental kelvon excitation frequency determine the inertial vortex mass \cite{Simula2018a}. Here we have further argued in favour of equality between the inertial and gravitational masses of the quantised vortices. The unified descripion of electromagnetism and gravity and the association of quantised vortices with massive charged particles, leads to the picture where the quantum Hall physics of a rapidly rotating neutral superfluid, condensation of elementary vortices into high density negative absolute temperature Onsager vortex clusters, and black hole thermodynamics with emergent quantum gravity are complementary ways to describe the states of dense vortex matter.

Recently, experiments on weakly interacting Bose--Einstein condensates have been used for simulating analogue spacetimes including inflatory cosmology \cite{Eckel2018a} and Hawking radiation \cite{Steinhauer2016a,Leonhardt2018a}, and it seems that quantum turbulent Bose--Einstein condensates \cite{Gauthier2019a,Johnstone2019a} may provide a fruitful analogue platform for further studies of emergent gravity, dark matter physics, and AdS/CFT correspondence \cite{Adams2012a}. Stretching the analogue in the opposite direction, it is amusing to contemplate the implications if the Universe were a superfluid hologram, gravity merely a manifestation of its quantum fluctuations, and the sought after dark matter just a terminal point of the photon dispersion relation---a Bose--Einstein condensate of ultra-weakly interacting photons---and that the seeming matter-antimatter asymmetry would be caused by evaporative heating induced negative temperature Onsager vortex clustering.  Finally, it is interesting to witness how modeling the dark matter in our Universe deploying quantum mechanical scalar fields is steadily making its way to mainstream cosmology \cite{GPEdm}.

\begin{acknowledgements}
This work was performed in part at Aspen Center for Physics, which is supported by National Science Foundation grant PHY-1607611. I am grateful for the Institute for Nuclear Theory at the University of Washington for its kind hospitality and stimulating research environment and the support by the INT's U.S. Department of Energy grant No. DE-FG02-00ER41132. I thank the Kavli Institute for Theoretical Sciences (KITS) at the University of Chinese Academy of Sciences (UCAS) for support and hospitality. Finally, I acknowledge financial support by Australian Research Council Discovery Projects Grant no. DP170104180 and Future Fellowships Grant no. FT180100020. \end{acknowledgements}

\bibliographystyle{apsrev}
\bibliography{gravirefs}

\end{document}